\documentclass[aps,preprint,amssymb,12pt,floatfix]{revtex4}
\usepackage{subfigure}
\usepackage{graphicx}
\usepackage{lscape,graphicx}
\usepackage{rotating}
\usepackage{color}
\usepackage{amsmath,amsthm}
\usepackage{wrapfig}

\usepackage{graphicx}
\usepackage{bm}

\def\prl#1#2#3{{ Phys.   Rev.   Lett.  } {\bf #1}, #2 (#3)}
\def\pla#1#2#3{Phys.   Lett.   A {\bf #1}, #2 (#3)}

\def\pre#1#2#3{Phys.   Rev.   E {\bf #1}, #2 (#3)}

\def\noi{\noindent}
\def\bc{\begin{center}}
\def\ec{\end{center}}
 \newcommand{\bea}{\begin{equation}}
 \newcommand{\eea}{\end{equation}\noi}
 \newcommand{\ber}{\begin{eqnarray}}
 \newcommand{\eer}{\end{eqnarray}\noi}
\begin{document}
\title{Exact solution of the Zwanzig-Lauritzen model of Polymer Crystallization under Tension}
\author{Himadri   S.   Samanta}
\author{D. Thirumalai}
\affiliation{Biophysics Program, Institute of Physical Science and Technology, 
University of Maryland, College Park, MD 20742}

\date{\today}
\begin{abstract}
We solve a two dimensional model for polymer chain folding in the presence of mechanical 
 pulling force ($f$)  exactly using equilibrium statistical mechanics.
Using analytically derived expression for the partition function we determine the phase diagram for the model in the $f$-temperature ($T$) plane.  A square root singularity in the susceptibility  indicates a 
 second order phase transition from a folded to an unfolded state at a critical force ($f_c$)
 in the thermodynamic limit of infinitely long polymer chain. Surprisingly,  the temperature dependence of $f_c$ shows a reentrant phase transition, which is reflected in an increase in $f_c$ as $T$  increases below a threshold value. For a range of $f$ values,  the unfolded state is stable at both low and high temperatures. The high temperature unfolded state is stabilized by entropy whereas the low temperature unfolded state is dominated by favorable energy. The exact calculation could serve as a bench mark for testing approximate theories that are used  in analyzing single molecule pulling experiments.      
\end{abstract}

\maketitle



Since the pioneering experiments \cite{GaubSCI97,Bustamante97Science,Tskhovrebova} demonstrating that titin, a large protein with multiple subunits,  can be unfolded upon application of mechanical force, $f$,  single molecule pulling experiments with increasing sophistication have been used to extract the folding landscapes of proteins \cite{stigler2011complex,Borgia08ARBiochem} and RNA \cite{Woodside08COCB,Hyeon05PNAS}. The experimentally accessible coordinate in these experiments is the molecular extension, $R$, which is conjugate to $f$ and represents typically the distance between the ends of the molecule of interest. Parameters describing the folding landscape, such as the barrier to unfolding at $f=0$, stability of the bound or folded states with respect to the unbound states, and location of the transition state are extracted from measurements (unbinding rates as a function of $f$ for example) by assuming that a single reaction coordinate, $R$, suffices. Although such measurements and interpretations have provided insights into some aspects of the folding mechanisms the inability to confirm the accuracy of the extracted parameters by using independent measurements of the key quantities remains a major, but not widely discussed, problem. This situation is exacerbated by a paucity of exactly solvable models (however see \cite{Barsegov08PRL}),  which can be used to test the accuracy of various commonly used approximate procedures  to extract  the folding landscape parameters from measured trajectories. 

Motivated in part for the reasons stated above, in this work we obtain an exact solution on the effect of force on a remarkable toy model introduced over forty years ago by Zwanzig and Lauritzen (ZL) \cite{Zwanzig68JCP} in the context of polymer crystallization. We construct the equilibrium phases of the ZL model in the [$f$,$T$] variables where $T$ is the temperature. Previously such phase diagrams have  been obtained numerically using models introduced to understand protein folding \cite{Klimov99PNAS,Klimov00PNAS} and also by theoretical methods for lattice homopolymer models \cite{Brak09JPhysA}.   As stated above, we were motivated to undertake this work in part to discover models for which exact solutions can be found so that the accuracy of approximate methods can be assessed. In addition, there is growing interest in the collapse transition of self attracting polymer chains, because of the relevance to protein folding \cite{Camacho93PNAS,Camacho93PRL,Schuler08COSB}.  Of course, chain folding is also at the heart of self-assembly of proteins and RNA. Such transitions are caused by a competition between the 
self attraction between segments of the polymer (tending to collapse the chain)  and the conformational  entropy that favors expansion. Typically, at low temperatures the collapsed state is stable 
dominated by energy or enthalpy.  Increase in $T$ populates  extended conformations that are stabilized by  entropy. Since, the ZL model includes both these effects we suspect that certain generic aspects of chain folding can also emerge from a deeper study of this interesting model. Although  the ZL model or the extension investigated here and elsewhere \cite{Caliri93PhysLettA,Lauritzen70JCP} may not be  directly applicable to proteins the study of such models is interesting in its own right, and for clarifying unusual aspects of phase transitions in polymers \cite{Klushin11JPhysA}.

 The ZL model was introduced as a caricature of 
polymer crystallization, which in the absence of $f$ exhibits  a phase 
transition that can be treated exactly by equilibrium statistical mechanics. The ZL  model leads to a novel second order-like phase transition
from an extended state to chain folded state in the thermodynamic limit of infinite chain length. The exact solution to the ZL model provided here shows folding-unfolding transition in the presence of mechanical force ($f$). In particular,
we show that a second order phase transition also occurs from  a folded state to an extended state at a critical force $f_c$ at a fixed $T$. The exact calculation of the  phase boundary separating the folded and extended states show a reentrant phase transition in which the extended state is stable at both high and low temperatures at finite $f$.

\section*{Zwanzig-Lauritzen model:} The ZL model considers a long polymer molecule whose conformations are  restricted to a plane. At $f=0$  the chain molecule folds into lamellar structures at low temperatures reminiscent of conformations adopted by  
crystalizing polymers.  A conformation of the ZL chain, displayed in Fig.~1, shows that it can fold upon itself by paying a pending penalty, which is compensated by attraction between the chain segments. Consequently, at $f=0$  the chain molecule folds into lamellar structures at low temperatures.   The molecule folds into
$N$ segments  ranging from one to infinity,
\bea\label{1}
1\le N<\infty.
\eea
The folds are uniform  occupying a length $q$ (Fig.~1).
Let the length of $i^{th}$ segment be $x_{i}$. It can have any length,
\bea\label{2}
0\le x_{i}<\infty,
\eea
with the constraint that the total length $L$, 
\bea\label{3}
\sum_{i=1}^{N}x_i+(N-1)q=L
\eea
be fixed.

We apply a mechanical force at one end of the molecule, say $\vec{f}$, with the other end  fixed.
The total energy of a conformation is associated with (1) a bending  energy $u$ for each fold (Fig.~1), (2) the interaction energy
between any pair of neighboring segments of length $x_{i}$ and $x_{i+1}$, and (3) stretching energy due mechanical 
force at one end. Two neighboring segments interact with short-ranged van der Waals forces. ZL assumed that the attractive  energy of interaction 
between $i_{th}$ and $(i+1)_{th}$ segments is $-J \ min(x_{i},x_{i+1})$,  where min($a,b$) stands for the smaller of the length of the two segments. 
The proportionality constant $J$ is taken to be positive. 
The stretching energy  due to the applied force $\vec{f}$ is taken to be,
\bea\label{4}
-\vec{f}\cdot (\vec{r})
\eea
where, $\vec{r}$ is the end-to-end distance of the molecule. In the present calculations we assume that the force is applied along the $x$ direction.
The total energy of a given conformation of the chain molecule is 
\ber\label{5}
E(x_1,....x_n)&=&(N-1)u -J\sum_{i=1}^{N-1}min(x_{i},x_{i+1})\\ \nonumber &&-f_{x}(N-1)q-f_y (x_N +p)
\eer
where $(N-1)q$ is the end-to-end distance along the $x$ direction and $x_N+p$ is the end to end distance along $y$
direction with arbitrary $p$, as shown in Fig.~1.

At first glance the ZL model resembles a one-dimensional model with short range interactions, which cannot exhibit a phase transition. However, ZL pointed out that this model is analogous to the familiar one dimensional Ising model with $N$ possible states for each of the $N$ sites. A two dimensional Ising model can be thought of as a one dimensional Ising system in which each site has $2^N$ sites. Thus, apparently the ZL model has enough of the two dimensional features  to exhibit an interesting second order transition from extended to folded states even in the absence of force.
 
\section*{Evaluation of the partition function in the presence of force:} We take the continuum limit of the model.   In this limit,  the canonical 
partition function $Z(L)$, which is  the sum of the Boltzmann factors $e^{-\beta E_N}$ of all 
configurations of a polymer molecule subject to the constraint of total length $L$, is
\ber\label{6}
Z(L)=\sum_{N=1}^{\infty}&& \int_{0}^{\infty} \cdots \int_{0}^{\infty} dx_1 \cdots dx_N \\ \nonumber &&
\delta(\sum_{i=1}^{N}x_i +(N-1)q-L) e^{-\beta E_N}
\eer
with $\beta=\frac{1}{K_BT}$, and $K_B$ is Boltzmann's constant.
As we sketch below the partition function $Z(L)$ can be evaluated analytically following the strategy developed by ZL. We start with  Eq.(\ref{6}) and perform a 
Laplace transform of the partition function with respect to length $L$ to eliminate the delta function on the 
right hand side of equation Eq.(\ref{6}). Then we invert the transform to obtain $Z(L)$. Using the following relation,
\bea
min(a,b)=\frac{1}{2}(a+b)-\frac{1}{2}|a-b|,
\eea
we can write the potential energy in the following form
\ber\label{8}
E_N=&&(N-1)u-\frac{1}{2}J\sum_{1}^{N-1} (x_i +x_{i+1})+\\ \nonumber &&
\frac{1}{2}J\sum_{1}^{N-1} |x_i -x_{i+1}|)-f_x(N-1)q-f_y(x_N+p).
\eer
We write Eq.(\ref{8})as,
\ber\label{9}
E_N=&& (N-1)u-J\sum_{1}^{N}x_i +\frac{1}{2}J(x_1+x_N \\ \nonumber &&
+\sum_{1}^{N-1}|x_i-x_{i+1}|)-f_x(N-1)q-f_y(x_N+p).
\eer
The sum $\sum_{1}^{N}x_i$ in equation (\ref{9}) can be replaced by $L-(N-1)q$, leading to
\ber\label{10}
E_N=&& (N-1)(u+Jq)-JL +\frac{1}{2}J(x_1+x_N \\ \nonumber &&
+\sum_{1}^{N-1}|x_i-x_{i+1}|)-f_x(N-1)q-f_y(x_N+p).
\eer
By substituting Eq.(\ref{10}) into Eq.(\ref{6}) we obtain,
\ber\label{11}
Z(L)=&&\sum_{N=1}^{\infty}[e^{-\beta(u+Jq-f_x q)}]^{N-1} e^{\beta JL}\\ \nonumber &&
\int_{0}^{\infty} \cdots \int_{0}^{\infty} dx_1 \cdots dx_N \delta(\sum_{i=1}^{N}x_i +(N-1)q-L)\\ \nonumber &&
\exp[-\frac{1}{2}\beta J(x_1 +x_N)+\frac{1}{2}J\sum_{1}^{N-1} |x_i -x_{i+1}|)]\\ \nonumber &&
e^{\beta f_y(x_N+p)}.
\eer
Taking the Laplace transform on both side to eliminate the delta function on the right hand side of
equation (\ref{11}), we obtain,
\ber\label{12}
&&\int_{0}^{\infty}dL e^{-(\eta+\beta J) L} Z(L)= \sum_{N=1}^{\infty}[e^{-\beta(u+Jq-f_x q)}]^{N-1}\\ \nonumber &&
e^{-\eta(N-1)q}\int_{0}^{\infty} \cdots \int_{0}^{\infty} dx_1 \cdots dx_N e^{-\eta \sum_{1}^{N}x_i}\\ \nonumber &&
\exp[-\frac{1}{2}\beta J(x_1 +x_N)+\frac{1}{2}J\sum_{1}^{N-1} |x_i -x_{i+1}|)]\\ \nonumber &&
e^{\beta f_y(x_N+p)}.
\eer

The integrals in Eq.(\ref{12}) are evaluated using the following method. We define a partial generating function
with the initial point fixed at the origin. Let us define a recursion relation of the partial generating function
with fixed $x_N = x$.
\bea\label{13}
g(z,x)=t e^{-\gamma |x-z|} e^{-\eta x} \left[\int_{0}^{\infty}g(x,y)+e^{-\gamma x}\right],
\eea
where $t=e^{-\beta(u+Jq-f_x q)}$ and $\gamma=\frac{1}{2} \beta J$. 
The needed generating function is
\bea\label{14}
G(t,\eta)=\int_{0}^{\infty} e^{\beta f_y x} g(z=0,x).
\eea
The term $\int_{0}^{\infty}g(x,y)$ in equation (\ref{13}) is the generating function in the absence of mechanical force, which is explicitly calculated in \cite{Zwanzig68JCP} by solving the differential equation given by,
\bea\label{15}
\int_{0}^{\infty}dy g(x,y)=-e^{-\gamma x}+\frac{\eta \alpha}{2t}\frac{J_{\nu}(\alpha e^{-\eta x/2})}{J_{\nu-1}(\alpha)},
\eea
with, $\alpha=(\frac{8\gamma t}{\eta^2})^{1/2}$ and $\nu=2\gamma/\eta$ and $u$ is replaced by $u+J q -f_x q$.
Using Eqs \ref{13}, \ref{14} and \ref{15}, we can evaluate the generating function in presence of mechanical force exactly.
We obtain
\ber\label{16}
&&G(t,\eta)=\frac{\Gamma(1-\frac{a}{2}+\nu)}{\Gamma(2-\frac{a}{2}+\nu)}\\ \nonumber &&\frac{\eta \alpha^{\nu+1}
{}_pF_q[\{1-\frac{a}{2}+\nu\};\{1+\nu,2-\frac{a}{2}+\nu\};-\frac{\alpha^2}{4}]}
{ J_{\nu-1}(\alpha) \Gamma(\nu+1)}.
\eer
with $a=\frac{2\beta f_y}{\eta}$ and $\eta>\beta f_y-2\gamma$.
Hence we get,
\bea\label{17}
\int_{0}^{\infty}dL e^{-(\eta+\beta J) L} Z(L)= e^{\beta(u+Jq-f_x q)}
e^{\eta q} G(t,\eta).
\eea
The partition function $Z(L)$ can be found by taking the inverse Laplace transform,
\bea\label{18}
Z(L)= e^{\beta(u+Jq-f_x q)}\frac{1}{(2\pi i)}
\int_{c-i \infty}^{c+i\infty}d\eta e^{\eta L+\eta q+\beta JL} G(t,\eta).
\eea
The contour of integration is a straight line, parallel to the imaginary axis, to the right of all singularities
of the integrand.

We now analyze the analytic properties of $G(t,\eta)$ in the complex $\eta$ plane. Originally, the function $G(\eta)$
is calculated for real $\eta$ with the imposed condition $\eta-\beta f_y+2\gamma >0$. Now, we wish to analytically continue 
$G(\eta)$ off the real $\eta$ axis. The Bessel function has branch points at the origin, and the usual cut along the negative 
real axis, but the related function, 
\bea\label{19}
\Lambda_{\nu-1}(z)=\Gamma(\nu)(\frac{z}{2})^{-\nu+1}J_{\nu-1}(z)
\eea
is an entire function of $z$ and a meromorphic function of order $\nu-1$. We start with real positive $\eta$ and hence real 
positive $\nu$, where the function $G$ is well defined, and can be written as,
\ber\label{20}
&& G(t,\eta)=\frac{\Gamma(\nu)\Gamma(1-\frac{a}{2}+\nu)}{\Gamma(\nu+1)\Gamma(2-\frac{a}{2}+\nu)}\\ \nonumber &&
\frac{\eta  \alpha^{2}{}_pF_q[\{1-\frac{a}{2}+\nu\};\{1+\nu,2-\frac{a}{2}\};\frac{-\alpha^2}{4}]}
{(2)^{1-\nu} \Lambda_{\nu-1}(\alpha)}
\eer

For simplicity, we assume that the pulling force is along $x$ direction (Fig.~1).
The function ${}_pF_q[\{1-\frac{a}{2}+\nu\},\{1+\nu,2-\frac{a}{2}\},\frac{-\alpha^2}{4}]$ reduces to $\Lambda_{\nu}$. 
Now $\Lambda_{\nu}$ and $\Lambda_{\nu-1}$ are analytic functions of 
complex $\nu$ everywhere except at singularities of the gamma function. At   these points, which are negative integers, they have simple pole. 
But, in the ratio, the singularities cancel out exactly at these points. This implies the singularities of $G$ in the complex plane are
determined solely by the zeros of the denominator, which are now poles. This accomplishes the desired analytic continuation
of the ratio of the positive real axis into the entire complex plane.

Let us define,
\bea\label{25}
\sigma=\frac{2}{(\beta J)^{1/2}}\exp[-\frac{\beta}{2}(u+Jq-f_x q)],
\eea
so that,
\bea\label{26}
\alpha=\sigma  \nu.
\eea
In order to evaluate the contour integral in Eq. \ref{18}, we need to evaluate the zeros of the denominator namely the zeros of the Bessel
function equation,
\bea\label{21}
J_{\nu-1}(\sigma  \nu)=0.
\eea
Each of the zeros will be function of $\sigma$ as well as the external force. To each zero, there corresponds to
a particular value of $\eta$, which are all real.
The sequence of all zeros of $J_{\nu-1}(\sigma  \nu)$, corresponds to the sequence of all real $\eta's$, 
which is given by, 
\bea\label{22}
\eta_1 >\eta_2 >\eta_3> \cdots.
\eea
indicating the largest one is $\eta_1$.
Let $R_i$ denote the residue at each pole of $G(\eta)$ corresponds to each $\eta_i$. We obtain,
\bea\label{23}
Z(L)=\exp(\beta(u+JL+Jq-f_x q))\sum_i R_i \exp(L\eta_i).
\eea

\section*{Phase Transition:} We are interested in the thermodynamic limit, corresponding to the length of the polymer molecule going to infinity.
In the large $L$ limit, the largest $\eta_1$ will dominate the sum. Consequently, the partition function
becomes,
\bea\label{24}
\lim_{L \rightarrow \infty}[L^{-1}\log Z(L)]=\beta J+\eta_1 (\sigma, \beta, f_x).
\eea
We have neglected terms that vanish in the large $L$ limit.
Thus, the evaluation of $Z(L)$ as well as the free energy reduces to the computation of a particular zero of the 
equation (\ref{21}), that gives the largest $\eta_1$.

The behavior of $\alpha$ for the largest order is known to be,
\bea\label{27}
\alpha=\nu-1+a_1 (\nu-1)^{1/3}
\eea
for large $\nu$,and
\bea\label{28}
\alpha= \sigma \nu.
\eea
Comparing these equations, we conclude that for $\sigma<1$, there is no solution for positive $\nu$.  There is an asymptotic solution $\nu=\infty$, and for $\sigma>1$, there is a unique 
solution $0<\nu<\infty$. We assume here that the fold energy $u$ is positive, $\sigma$ is monotonically increasing function of both temperature and force.
Thus, $\sigma$  is unity at the transition point determined by either
\bea\label{29}
\frac{2}{(\beta_c J)^{\frac{1}{2}}} \exp[-\frac{\beta_c}{2} (u+Jq-f_x q)]=1,
\eea
for fixed $f_x$, or
\bea\label{30}
\frac{2}{(\beta J)^{\frac{1}{2}}} \exp[-\frac{\beta}{2} (u+Jq-f_c q)]=1,
\eea
for arbitrary $\beta$.
Equation (\ref{30}) represents the exact form of the critical force for arbitrary $\beta$.

In the limit $\beta \rightarrow \beta_c$ from above or $f_x \rightarrow f_c$ from above at a constant $\beta$, $\nu$ approaches $\infty$, so that,
\bea\label{31}
\sigma \nu =\nu+a \nu^{1/3}.
\eea
It follows that
\bea\label{32}
\nu=a^{3/2} (\sigma-1)^{-3/2},     
\eea
if $T \sim T_{c}^{+}$ or $f_x \sim f_{c}^{+}$.
or equivalently,
\bea\label{33}
\nu^{-1}=a^{-3/2} (\sigma-1)^{3/2}.     
\eea

\section*{Thermodynamics} We discuss the thermodynamic properties of the model in the long chain limit. Let $A$ denote the Helmholtz free
energy of the system. In the thermodynamic limit, the free energy per unit length is given by,
\bea\label{34}
\lim_{L\rightarrow \infty}(\frac{A}{L})=-J-\beta^{-1} \eta_1.
\eea
By defining the reduced temperature $T'= \frac{1}{\beta J}$, we obtain the reduced free energy,
\bea\label{35}
A'=\lim_{L\rightarrow \infty}(\frac{A}{LJ})=-1-T' \eta_1.
\eea
The reduced mean energy is given by,
\bea\label{35a}
E'=\lim_{L\rightarrow \infty}(\frac{E}{LJ})=-T'^{2} \frac{\partial A'/T'}{\partial T'}.
\eea
The average end-to-end distance in the direction of $f_x$ is $<(N-1)q>$ can be found by differentiating 
the partition function with respect to $f_x$,
\bea\label{36}
<(N-1)q>= \beta^{-1} \frac{\partial \ln Z}{\partial f_x}|_{T'}.
\eea

In the thermodynamic limit, the average end-to-end distance per unit length, $l$, along the $x$ direction is,
\bea\label{37}
l=\lim_{L\rightarrow \infty}(\frac{1}{L}<(N-1)q>)=-J (\frac{\partial A'}{\partial f})_{T'}.
\eea
We are interested in the vicinity of the transition point, which is in the neighborhood of $\sigma=1$.
Suppose that, $\sigma=1$, when $T'=T_c(\gamma)$ when $f_x$ is fixed, Taylor series expansion of 
$\sigma$ is given by,
\bea\label{38}
\sigma(T')=1+[\frac{1}{2T_c}+\frac{u/J+q}{2T_{c}^{2}}-\frac{f_x q}{2 T_{c}^{2}J}](T'-T_c)+\cdots
\eea
Similarly, for arbitrary $T'$, Taylor series expansion of $\sigma$ around the critical force $f_c$ is 
given by,
\bea\label{39}
\sigma=1+[\frac{q}{2 T' J}](f_x-f_c)+\cdots
\eea

First consider the case with $ f_x= 0$.  The expansion in (\ref{38}) shows that $\nu^{-1}$ is proportional to three halves power of the temperature deviation from $T_c$.
Now, for $T'<T_c$, $\sigma$ is less than unity and $\eta_1=0$. When $T'$ becomes $T_{c}^{+}$, $\sigma$ becomes greater than unity.
The free energy is given by,
\bea\label{40}
A'=-1-a_{1}^{-3/2} [\frac{1}{2T_c}+\frac{u/J+q}{2T_{c}^{2}}]^{3/2}(T'-T_c)^{3/2}
\eea
if $T'>T_c$, and,
\bea\label{41}
A'=-1
\eea
if $T'<T_c$.

Similarly, for specific heat,
\bea\label{42}
(\frac{\partial E}{\partial T'})_{\gamma}=0
\eea
if $T'<T_c$, and 
\bea\label{43}
(\frac{\partial E}{\partial T'})_{\gamma}=\frac{3}{4} [T_c][\frac{1}{2T_c}+\frac{u/J+q}{2T_{c}^{2}}]^{3/2}(T'-T_c)^{-1/2}
\eea
if $T'>T_c$.
Specific heat shows  inverse square singularity at the transition temperature. The discontinuity in specific heat shows a second 
order phase transition at $T'=T_c$ for fixed $f_x$ =0.

It can be easily seen that in the absence of force, and for negative value of the bending energy $u$, in the interval
$-\frac{J}{4e}-Jq<u<-Jq$, $\sigma$ is infinite for both $T'=0$ and $T'=\infty$, with  a single minimum below unity at 
a finite temperature. It follows that $\sigma$ will intersect unity at two temperatures and hence there are two phase transitions. 
At low enough temperatures, with negative bending energy, increasing the number of folds will minimize the total energy. This leads to the stabilization of the stable folded state. Such a possibility is not realized for positive fold energy.

Now, we  fix $T$  and vary the applied mechanical force.  The expansion in (\ref{39}) shows that $\nu^{-1}$ is proportional to the three halves power of the 
force deviation from $f_c$. For $f_x<f_c$, $\sigma$ is less than unity and $\eta_1 =0$.  As $f_x \rightarrow f_{c}^{+}$,
$\sigma$ becomes greater than unity.
The free energy is given by
\bea\label{44}
A'=-1
\eea
if $f_x<f_c$, and
\bea\label{45}
A'=-1-[\frac{q}{2 a_1 T' J}]^{3/2}(f_x-f_c)^{3/2}
\eea
if $f_x>f_c$.

The average end to end distance $l$ per unit length in the thermodynamic limit is found in (\ref{37}).
If $\sigma<1$, then we have $\eta_1 =0$ and the mean distance per unit length is zero. In the thermodynamic 
limit, any finite end-to-end distance corresponds to $l=0$. The average end-to-end distance per unit length
vanishes when $f_x<f_c$ corresponding to the folded state, and increases with the square root of the force difference above $f_c$ for arbitrary
$T'$ representing the unfolded state as follows,
\bea\label{46}
l=0 \ \ \ \ \ \ \ \ \ \ \ \, f_x<f_c,
\eea
and
\bea\label{47}
l=\frac{3}{2}[\frac{q}{2 a_1 T' J}]^{3/2}(f_x-f_c)^{1/2},
\eea
if, $f_x>f_c$.
Thus, the folded-unfolded transition in this model is second order even in the presence of force. Our calculations further reveal that
the susceptibility $\chi=\frac{\partial l}{\partial f_x}$ of the polymer diverges as
\bea\label{48}
\chi\propto (f_x-f_c)^{-1/2}
\eea
as $f_x$ approaches the transition point from above. The divergence in susceptibility at transition point
shows the second order phase transition with the characteristic critical exponent $1/2$.

\section*{Phase diagram in the $[f,T]$ plane:} In Fig.~2, we show the regions of $f-T$ plane, where the  chain is extended (white) and
where it is chain folded (shaded) for this model. Remarkably, the phase diagram exhibits a  reentrant behavior.  The boundary separating the two phases occurs when
\bea\label{49}
2 T'^{1/2} e^{-\frac{\frac{u}{J}+q}{2T'}} e^{\frac{f_c q}{2JT'}}=1.
\eea
This is an exact form for the temperature dependence of the critical force, $f_c$. The transition curve separates the region of extended chain and chain folded
configurations. Upon crossing the phase boundary from below the folded chain undergoes a transition to the extended state  via second order transition.
From Eq.(\ref{49}), we can write,
\bea\label{50}
\frac{f_c}{J} q = (\frac{u}{J}+ q)-T' \log(4) -T' \log(T').
\eea

The transition line in Fig.~2  increases at low temperatures due to the leading term $T' \log T'$ in Eq.(\ref{50}).
 In the interval $u+Jq<f_c q<u+Jq+\frac{J}{4e}$, the folding transition occurs at a critical force over a range of  temperatures.  However, upon 
further lowering the temperature the chain becomes extended a process that is reminiscent of 'cold denaturation'. Hence, the extended state is stable
at both high and low temperatures, implying that there is reentrant phase transition. At low enough temperatures, for the case of 
$f_c q >u+Jq$, large end-to-end distance  minimizes the total energy corresponding to a stable extended state.   
 In particular, Fig.~2 shows the unusual behavior that at low enough temperatures the critical force required to unfold the chain is less that at a higher temperature. For example, $f_c$ at $T = 0.05$ is less than $f_c$ at $T = 0.1$.

\section*{Conclusions:} The exact solution of the ZL model for polymer chain folding in the presence of force shows  a second order ordering transition from a folded to an unfolded state provided the values of force are less than a critical value, $f_c(T)$. This prediction is not in accord with mean field theory, which suggests that  the unfolding transition in the presence of external mechanical force is likely to be 
first order in all dimensions \cite{Halperin91EPL}. Indeed, molecular simulations of forced-unfolding of proteins  exhibits a  first order phase transition in three dimensions \cite{Klimov99PNAS,Klimov00PNAS}.
However in two dimensions, extensive Monte Carlo simulations performed on a self avoiding walk model in poor solvent \cite{Grassberger02PRE}
suggests that the folded to unfolded transition could have the hall marks of a  second order, in accord with the present calculations.   Furthermore, scaling analysis  shows a second 
order phase transition at a critical force $f_c$ \cite{Marenduzzo03PRL,RosaMarenduzzo}. Renormalization group analysis of polymer unfolding in special lattices also reveals a change in the nature 
of the phase transition as the spatial dimension exceeds two \cite{Vidanovic11EuroPhysJB}. Thus, the upper critical dimension for force-induced unfolding of self-attracting polymers and hence proteins is likely to be three.

Interestingly, for a range of $f < f_c(T)$ the ZL model exhibits a reentrant phase transition in which the disordered state is stable at low and high temperatures. As a consequence,  $f_c(T)$ increases as $T$ increases in the low $T$ regime. Force-induced reentrant behavior may be a common feature in other model polymeric systems \cite{Osborn10JStatPhys,Skvortsov12PRE,Marenduzzo03PRL}. A recent simulation study in the constant force ensemble \cite{Skvortsov12PRE} showed that the temperature-dependent force-induced desorption of two-dimensional self-avoiding polymer  is very similar in shape to the curve plotted in Fig.~2.  The generality of reentrant phase transitions induced by force, which is similar to cold denaturation in proteins, remains to established. It would be of particular interest to characterize such transitions, if they exist, in proteins. This would require performing pulling experiments as a function of temperature and force. 

In addition to adding to the collection of exactly solvable polymer models exhibiting phase transitions,  the present work could be used to test approximate theories  used to obtain folding thermodynamics of proteins using single molecule pulling experiments, which assume that extension is an excellent parameter.

{\bf Acknowledgements:} DT is grateful to Robert Zwanzig for introducing him to the ZL model over twenty five years ago.  This work was supported by a grant from the National Institutes of Health (No. GM089685) and the National Science Foundation (CHE 09-14033).


\begin{thebibliography}{99}
\bibitem{GaubSCI97} M. Rief and H. Gautel and F. Oesterhelt and J. M. Fernandez and H. E. Gaub, Science {\bf 276}, 1109 (1997).
\bibitem{Bustamante97Science} M. Z. Kellermeyer and S. B. Smith and H. L. Granzier and C. Bustamante, Science {\bf 276}, 1112 (1997).
\bibitem{Tskhovrebov} L. Tskhovrebova and J. Trinick and J.A. Sleep and R.M. Simmons, Nature {\bf 387}, 308 (1997).
\bibitem{stigler2011complex} Stigler, J. and Ziegler, F. and Gieseke, A. and Gebhardt, J.C.M. and Rief, M., Science {\bf 334}, 512 (2011).
\bibitem{Borgia08ARBiochem} Borgia, A. and Williams, P.M. and Clarke, J., Annu. Rev. Biochem. {\bf 77}, 101 (2008), ISSN 0066-4154.
\bibitem{Woodside08COCB} M. T. Woodside and C. Garcia-Garcia and S. M. Block, Curr. Opin. Chem. Biol. {\bf 12}, 640 (2008).
\bibitem{Hyeon05PNAS} C. Hyeon and D. Thirumalai, Proc. Natl. Acad. Sci. USA {\bf 102}, 6789 (2005).
\bibitem{Barsegov08PRL} V. Barsegov and G. Morrison and D. Thirumalai, \prl{100}{248102}{2008}.
\bibitem{Zwanzig68JCP} Zwanzig, R and Lauritzen, J. I., J. Chem. Phys. {\bf 48}, 3351 (1968)
\bibitem{Klimov99PNAS} D. K. Klimov and D. Thirumalai, Proc. Natl. Acad. Sci. USA {\bf 96}, 6166 (1999).
\bibitem{Klimov00PNAS} D. K. Klimov and D. Thirumalai, Proc. Natl. Acad. Sci. USA {\bf 97}, 7254 (2000).
\bibitem{Brak09JPhysA} Brak, R. and Dyke, P. and Lee, J. and Owczarek, A. L. and Prellberg, T.
   and Rechnitzer, A. and Whittington, S. G., J. Phys. A-Mathematical and Theoretical {\bf 42}, 085001 (2009).
\bibitem{Camacho93PNAS} C. J. Camacho and D. Thirumalai, Proc. Natl. Acad. Sci. USA {\bf 90}, 6369 (1993).
\bibitem{Camacho93PRL} Camacho, C. J. and Thirumalai, D., \prl{71}{2505}{1993}.
\bibitem{Schuler08COSB} B. Schuler and W. A. Eaton, Curr. Opin. Struct. Biol. {\bf 18}, 16 (2008).
\bibitem{Caliri93PhysLettA} Caliri, A and Bohr, H and Wolynes, P., \pla{183}{327}{1993}. 
\bibitem{Lauritzen70JCP} Lauritzen, J.I. and Zwanzig, R., J. Chem. Phys. {\bf 52}, 3740 (1970).
\bibitem{Klushin11JPhysA} Klushin, L. I. and Skvortsov, A. M., J. Phys. A-Mathematical and Theoretical {\bf 44}, 473001 (2011).
\bibitem{Halperin91EPL} Halperin, A and Zhulina, E.B., Europhys. Lett. {\bf 15}, 417 (1991).
\bibitem{Grassberger02PRE} Grassberger, P and Hsu, H.P, \pre{65}{031807}{2002}
\bibitem{Marenduzzo03PRL} Marenduzzo, D and Maritan, A and Rosa, A and Seno, F., \prl{90}{088301}{2003}.
\bibitem{RosaMarenduzzo} Rosa, A and Marenduzzo, D and Maritan, A and Seno, F., \pre{67}{041802}{2003}. 
\bibitem{Vidanovic11EuroPhysJB} Vidanovic, I. and Arsenijevic, S. and Elezovic-Hadzic, S., Europhys. J. B pp. 291-302 (2011).
\bibitem{Osborn10JStatPhys} Osborn, J. and Prellberg, T., J. Stat. Phys. p. P09018 (2010).
\bibitem{Skvortsov12PRE} Skvortsov, A. M. and Klushin, L. I. and Polotsky, A. A.
   and Binder, K., \pre{85}{031803}{2012}.
\end{thebibliography}

\newpage
\begin{center}
\textbf{\large{Figure Captions}}
\end{center}

Fig.~1: Conformation of a polymer chain in the ZL model. Two contiguous vertical lines define a fold. There is a energy gain per each fold that is achieved by paying a bending penalty. The length of each fold is $q$. A force $f$ is applied to one end of the chain. The calculations are done by assuming that the force is along the $x$ axis.

Fig.~2: Phase diagram in the $f-T$ plane with $q=2$ and $u/J=0.01$. At low temperatures the critical force increases as $T$ increases reflecting the observed reentrant behavior.
\newpage

\begin{figure}[h]
\includegraphics[scale=0.800]{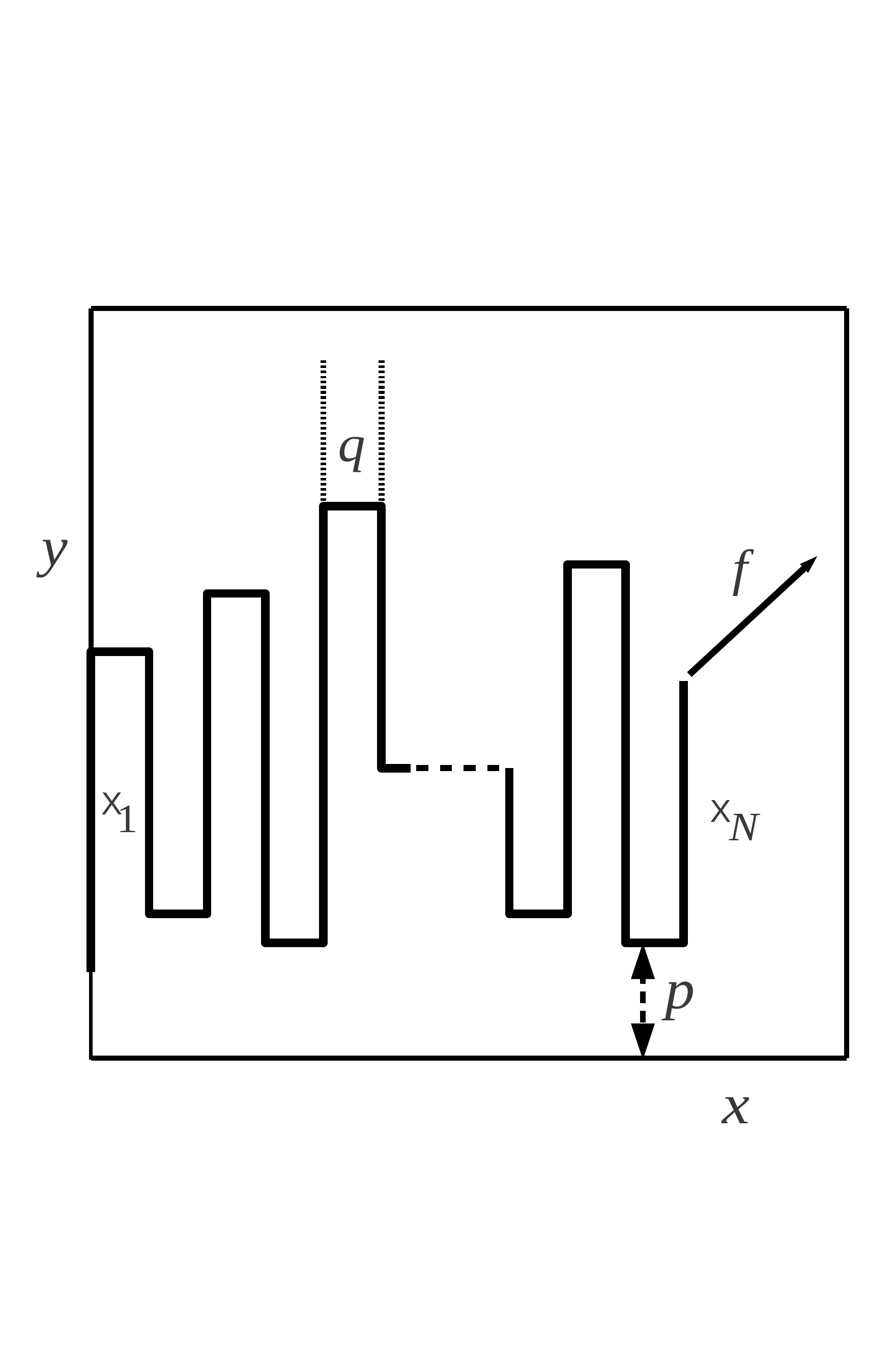}
\end{figure}

\newpage
\begin{figure}[h]
\begin{center}
\includegraphics[scale=0.800]{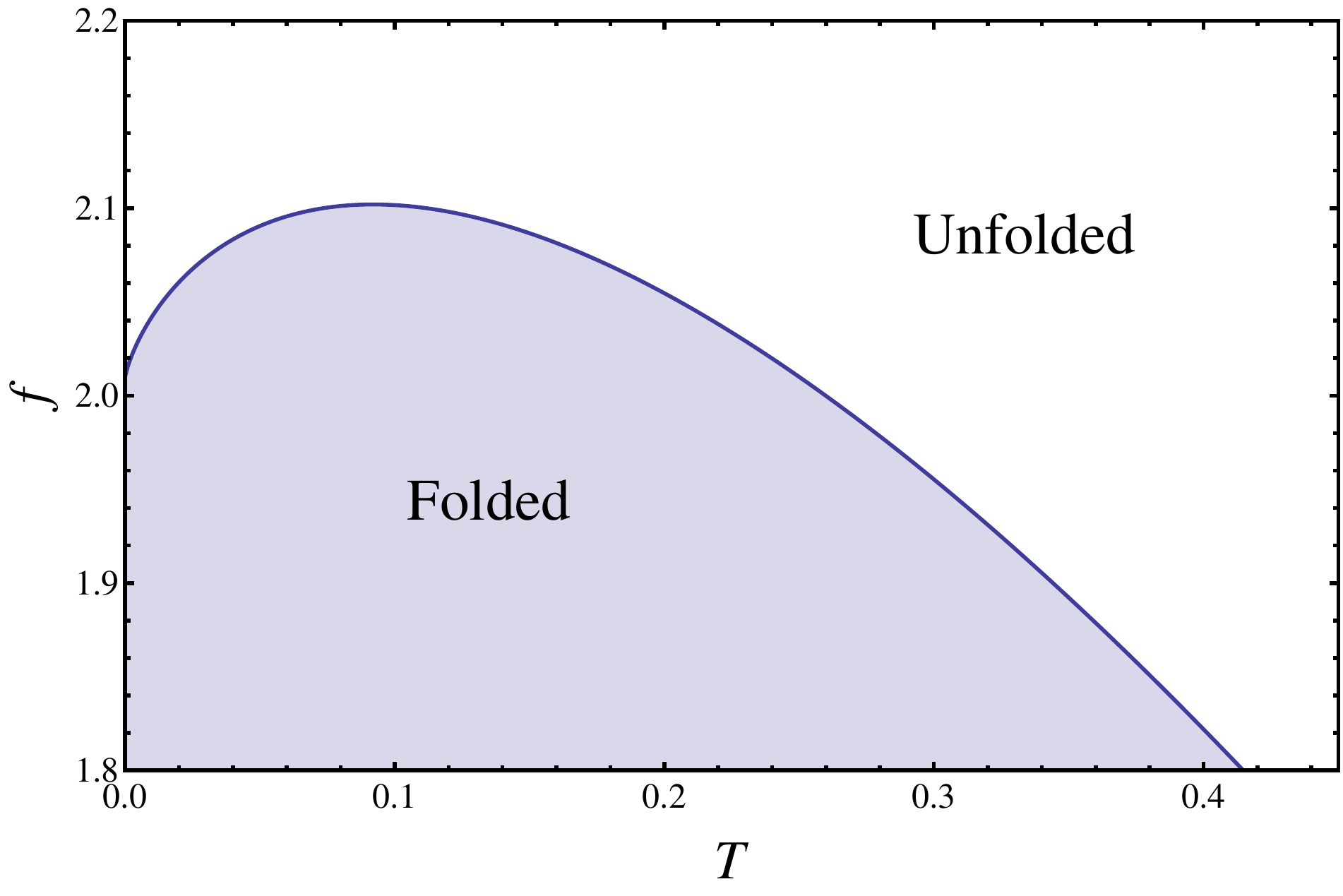}
\end{center}
\end{figure}

\end{document}